\documentclass[%
 reprint,
 amsmath,amssymb,
 aps,
]{revtex4-1}

\usepackage{graphicx}
\usepackage{dcolumn}
\usepackage{bm}
\usepackage{comment}
\usepackage{color}
\usepackage{float}
\usepackage{cancel}

\newcommand{\Hop}{\hat{H}}
\newcommand{\aop}{\hat{a}}
\newcommand{\aopd}{\hat{a}^\dagger}
\newcommand{\bop}{\hat{b}}
\newcommand{\bopd}{\hat{b}^\dagger}

\newcommand{\alphain}{\alpha_{\mathrm{in}}}
\newcommand{\alphaout}{\alpha_{\mathrm{out}}}

\begin{document}

\preprint{APS/123-QED}

\title{Coupling a superconducting quantum circuit to a phononic crystal defect cavity}

\author{Patricio Arrangoiz-Arriola}
\author{E. Alex Wollack}
\author{Marek Pechal}
\author{Jeremy D. Witmer}
\author{Jeff T. Hill}
\author{Amir H. Safavi-Naeini}
 \affiliation{Department of Applied Physics and Ginzton Laboratory, Stanford University \\
 				348 Via Pueblo Mall, Stanford CA 94305 USA}

\date{\today}

\begin{abstract}

 Connecting nanoscale mechanical resonators to microwave quantum circuits opens new avenues for storing, processing, and transmitting quantum information. In this work, we couple a phononic crystal cavity to a tunable superconducting quantum circuit. By fabricating a one-dimensional periodic pattern in a thin film of lithium niobate and introducing a defect in this artificial lattice, we localize a 6 gigahertz acoustic resonance to a wavelength-scale volume of less than one cubic micron. The strong piezoelectricity of lithium niobate efficiently couples the localized vibrations to the electric field of a widely tunable high-impedance Josephson junction array resonator. We measure a direct phonon-photon coupling rate $g/2\pi \approx 1.6 \, \mathrm{MHz}$ and a mechanical quality factor $Q_\mathrm{m} \approx 3 \times 10^4$ leading to a cooperativity $C\sim 4$ when the two modes are tuned into resonance. Our work has direct application to engineering hybrid quantum systems for microwave-to-optical conversion as well as emerging architectures for quantum information processing.

\end{abstract}

\maketitle

\section{Introduction}

Compact and low-loss {acoustic}  wave devices that perform complex signal processing at radio frequencies are ubiquitous in classical communication systems~\cite{Hashimoto2009}. Much like their classical counterparts, emerging quantum machines operating at microwave frequencies~\cite{Devoret2013} also stand to benefit from their integration with these devices. This is conditioned on the realization of sufficiently versatile \emph{quantum phononic} technologies. Several promising approaches have emerged in the last few years. Each has its own strengths and weaknesses and they can be broadly categorized by the degree to which the acoustic  waves are confined as compared to their wavelength. In a series of remarkable experiments, thin-film~\cite{OConnell2010}, surface~\cite{Gustafsson2014a,  Manenti2017, Moores2017, Noguchi2017}, and bulk acoustic wave resonators~\cite{Chu2017, Kervinen2018} made of piezoelectric materials have coupled gigahertz phonons with varying levels of confinement to superconducting circuits.  Nonetheless, smaller mode volumes, lower losses, and greater control over the mode structure are desired. 

One of the most promising approaches for realizing ultra-low-loss mechanical resonators is to use phononic crystal cavities that confine acoustic waves in all three dimensions. Wavelength-scale confinement and periodicity qualitatively alter the properties of waves and allow far greater control over the phonon density of states. Periodic patterning of a thin slab of elastic material can give rise to a phononic bandgap --  a range of frequencies devoid of propagating waves. By introducing defects in such a crystal, mechanical energy is localized at the wavelength scale \cite{Eichenfield2009,Alegre2010,Bochmann2013, Safavi-Naeini2014, Balram2015}  without any ``clamping'' losses. The existence of a full phononic bandgap eliminates all modes into which a phonon can be linearly scattered, leading to a significant increase in the coherence time of such resonators. For example, lifetimes on the order of one second corresponding to $Q>10^{10}$ have been optically measured in $5~$GHz phononic crystal cavities made from silicon~\cite{Maccabe2018APS}. Moreover, the small mode volume of phononic crystal cavities leads to a dramatic reduction in the density of spurious modes that can negatively impact performance of quantum acoustic systems, while enabling a denser packing of devices for greater scalability.

The greater confinement and control over the acoustic mode structure comes at the cost of weaker coupling. At gigahertz frequencies, the modes of phononic crystal cavities are confined to extremely small volumes ($\lesssim \, 1 \mu\mathrm{m}^3$). This leads to smaller forces for a given oscillating voltage when compared to approaches with transducer dimensions of tens to hundreds of microns~\cite{Arrangoiz-Arriola2016}. 
Up to now, it has only been possible to efficiently read out and couple to localized modes of phononic crystal cavities with optical photons where the electromagnetic energy is similarly localized~\cite{Eichenfield2009,Liang2017}. Nonetheless, to connect these systems to microwave superconducting quantum circuits, efficient and tunable coupling between microwave photons and phonons is needed. In this work we demonstrate the direct coupling of a superconducting circuit to a wavelength-scale phononic nanocavity, opening a new avenue in quantum acoustics.

\begin{figure}[ht]
    \centering
    \includegraphics[width=0.8\linewidth]{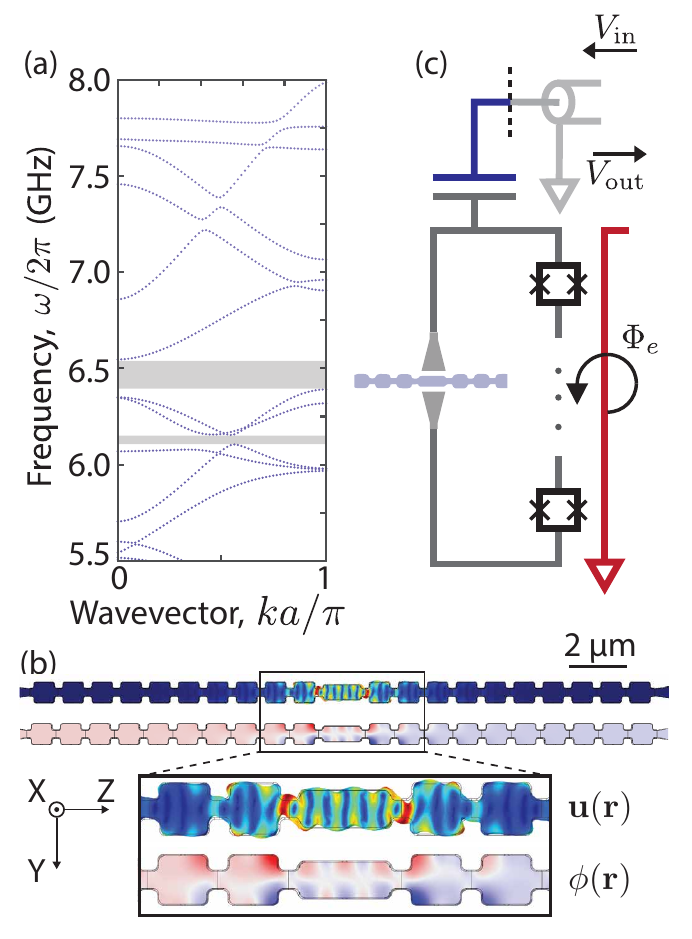}
    \caption{Concept and design. (a) Phononic bands of a $\mathrm{LiNbO_3}$ quasi-one-dimensional phononic crystal with lattice constant $a = 1 \, \mu\mathrm{m}$, showing the bands of all possible mode polarizations in the range of frequencies relevant to this work. A complete bandgap near $\nu = 6.5 \, \mathrm{GHz}$ is clearly visible, with a narrower gap also visible below. Other relevant simulation parameters (matching those of the fabricated structures) are the length and width of the connecting struts (320 nm and 240 nm, respectively), the film thickness (224 nm), and the sidewall angle ($5^\circ$). (b) Deformation $\mathbf{u}(\mathbf{r})$ and electrostatic potential $\phi(\mathbf{r})$ of a mode localized at the defect site, at frequency $\nu = 6.48 \, \mathrm{GHz}$ near the center of the bandgap. Modes of this polarization can be coupled to electric fields pointing in the direction perpendicular to the crystal lattice. Here the length and width of the defect are $a_\mathrm{def} = 1.6 \, \mu \mathrm{m}$ and $w_\mathrm{def} = 500 \, \mathrm{nm}$, respectively. (c) Schematic of the device, including the drive/readout line (blue) capacitively coupled to the resonator, the flux line (red) used to flux bias the SQUID array, and the electrodes (gray) that couple the circuit to the phononic cavity (light blue). The $\mathrm{LiNbO_3}$ crystal axes are indicated.}
    \label{fig:concept}
\end{figure}

\begin{figure*}[ht]
    \centering
    \includegraphics[width=\linewidth]{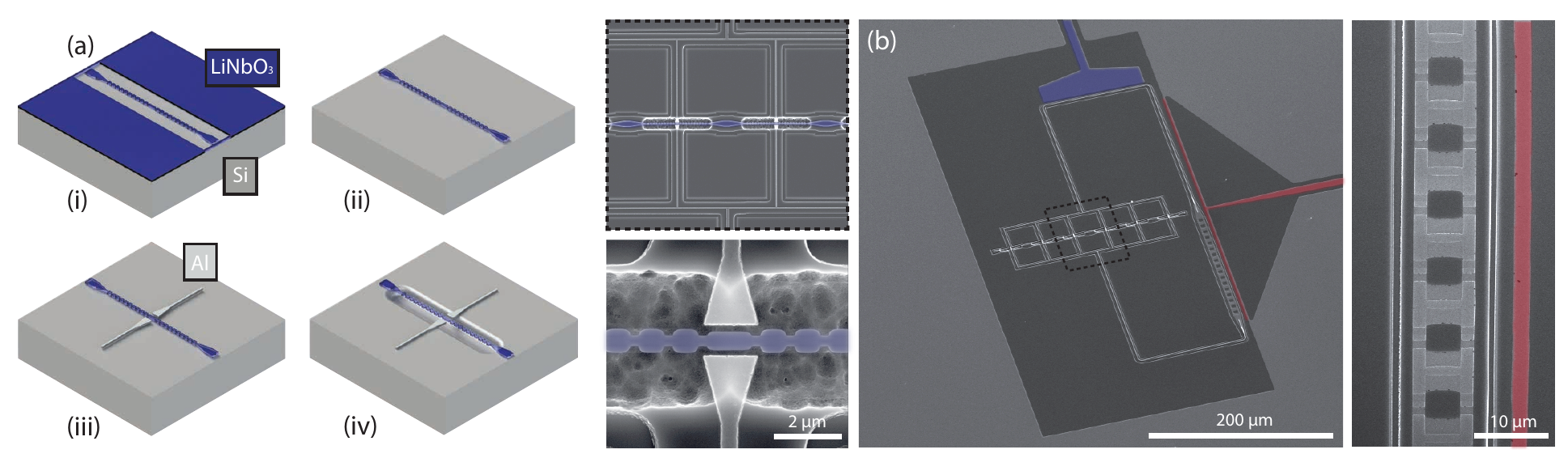}
    \caption{Device fabrication. (a) Schematic of the fabrication process, including (i) electron-beam patterning and argon milling of the phononic nanostructures in the $\mathrm{LiNbO_3}$ film, (ii) masked removal of the film from the rest of the substrate, (iii) deposition of all metallization layers, and (iv) masked undercut of the structures. The last step suspends both the phononic cavities and the edge of the coupling electrodes over an etched Si trench. (b) False-colored scanning-electron micrographs of the final device. The charge and flux lines are highlighted in blue and red, respectively. A close-up of the SQUID array clearly shows the Al/AlOx/Al junctions and the trenching in the Si substrate on either side of the array, produced by a deliberate gap between the electron-beam and photolitography masks used to pattern the $\mathrm{LiNbO_3}$ film which results in the Si getting etched twice in those regions. To the left of the SQUID array a group of 6 phononic crystal cavities, coupled to the array by 200 nm thick Al wires, is visible and highlighted by dashed black lines. A close-up of this region shows the $\mathrm{LiNbO_3}$ structures (highlighted blue), including the bandgap regions, the defect site surrounded by the partially suspended aluminum electrodes, and the etched Si trench. The porous-like surface of the etched Si is attributed to micromasking during the $\mathrm{XeF_2}$ undercut, but is likely unimportant as it is located far ($>2 \mu\mathrm{m}$) from the region between the electrodes, where the electric fields are strongest.}
    \label{fig:fab}
\end{figure*}

\section{Device design and fabrication}

At the heart of our device lies a suspended quasi-one-dimensional phononic crystal fabricated from a 200-nanometer-thick film of lithium niobate ($\mathrm{LiNbO}_3$). The crystal has a lattice constant of $1\,\mu\mathrm{m}$ and has a complete phononic bandgap in the vicinity of $\nu = 6 \, \mathrm{GHz}$ [Fig. \ref{fig:concept}(a)]. This bandgap is used to localize the resonances of a single defect site introduced in the center of the lattice. In particular, we engineer a defect mode with a strain field $S$ that generates a charge polarization $P_i = e_{ijk}S_{jk}$ that is predominantly aligned in-plane in the direction perpendicular to the lattice [Fig. \ref{fig:concept}(b)]; here $e$ is the piezoelectric coupling tensor of $\mathrm{LiNbO_3}$. We then use this polarization to couple the defect mode to the microwave-frequency electric field of a readout circuit, applied by gate electrodes placed within $200$ nanometers of the defect. The readout circuit is a lumped-element microwave resonator formed from the capacitance $C_\mathrm{r}$ of the gate electrodes and a series of Josephson junctions in a superconducting quantum interference device (SQUID) array configuration with total Josephson inductance $L_\mathrm{r} = \Phi_0^2/E_\mathrm{J}(\Phi_\mathrm{e})$ [Fig. \ref{fig:concept}(c)], where $\Phi_0 = \hbar/2e$ is the reduced flux quantum. The effective Josephson energy $E_\mathrm{J}(\Phi_0)$ of the array depends on the external flux $\Phi_\mathrm{e}$ threading the SQUIDs, making the resonator frequency $\omega_\mathrm{r} = 1/\sqrt{L_\mathrm{r} C_\mathrm{r}}$ tunable by applying a small current to an on-chip flux line. In addition, the small parasitic capacitance of the array enables us to achieve a relatively large resonator impedance $Z_\mathrm{r} = \sqrt{L_\mathrm{r}/C_\mathrm{r}} \approx 580 \,\Omega$. This is an important feature of our  device, as the piezoelectric coupling strength is proportional to the zero-point voltage fluctuations of the circuit and $V_\mathrm{zp} \sim \sqrt{Z_\mathrm{r}}$. The resonator impedance is largely limited by the presence of the flux line (highlighted in red in Fig.~\ref{fig:fab}), which is a major source of parasitic capacitance between the two nodes of the resonator. 

Thin-film $\mathrm{LiNbO_3}$ has recently gained prominence in the realm of classical radio frequency systems~\cite{Wang2013a,Pop2017,Vidal-alvarez2017}. Here we perform the device fabrication on a $500\,\mathrm{nm}$ film of X-cut $\mathrm{LiNbO_3}$ on a $500\,\mu\mathrm{m}$ high-resistivity ($>3\,\mathrm{k}\Omega\cdot \mathrm{cm}$) Si substrate and involves seven masks of lithography consisting of the following four stages [see Fig. \ref{fig:fab}(a)]: 1) $\mathrm{LiNbO_3}$ film thinning, 2) patterning of phononic nanostructures, 3) deposition of Al layers, including all microwave circuitry and Josephson junctions, and 4) masked undercut of structures. The film is first thinned down to the target thickness (approximately 200 nm for this device) by blanket argon milling. We then expose a mask on positive resist with a single step of electron-beam (ebeam) lithography and transfer it to the $\mathrm{LiNbO_3}$ film with an optimized argon milling process to form the phononic nanostructures. Now masking only the structures, an argon milling step is done to remove the $\mathrm{LiNbO_3}$ film from the entire sample. This step allows us to place all microwave circuits on a high-resistivity silicon substrate where they are not vulnerable to acoustic radiation losses induced by the piezoelectric film. Aluminum microwave ground planes and feedlines are defined on the exposed silicon via liftoff, and the SQUID arrays are fabricated with a Dolan bridge double-angle evaporation process to grow the Al/AlOx/Al junctions \cite{Dolan1977,Stockklauser2017}. The gate electrodes used to address the phononic defect sites are patterned with a separate ebeam mask and normal-incidence Al evaporation, and finally a bandage process \cite{Dunsworth2017} is used to ensure lossless superconducting connections between all metalization layers. As a final step, we release the structures with a masked $\mathrm{XeF}_2$ dry etch that etches the underlying Si with extremely high selectivity to the $\mathrm{LiNbO_3}$ and the Al~\cite{Pop2017,Vidal-alvarez2017}, leaving all aluminum layers intact at the end of the process.

\begin{figure*}[ht]
    \centering
    \includegraphics[width=0.9\linewidth]{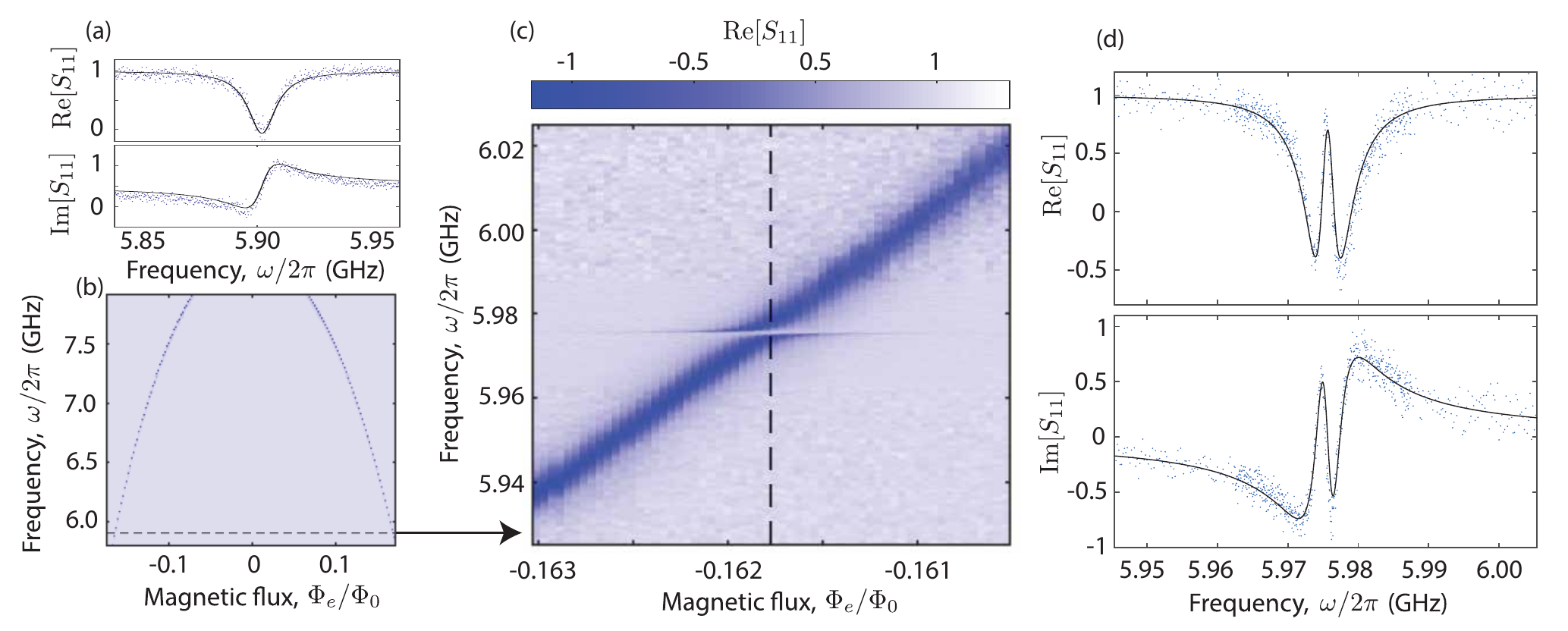}
    \caption{Linear spectroscopy. (a) Reflection spectrum $S_{11}(\omega)$ of the SQUID-array resonator tuned to a frequency of $\omega_\mathrm{r}/2\pi = 5.90 \, \mathrm{GHz}$, including the raw data and the fit to the model (solid lines). The data is normalized to a spectrum collected with a very large probe power ($P = 0 \, \mathrm{dBm}$), where the nonlinear resonance is saturated and absent from the spectrum. At this frequency, we obtain resonator decay rates $\kappa/2\pi = 11 \, \mathrm {MHz}$ and $\kappa_\mathrm{e}/2\pi = 6.3 \, \mathrm{MHz}$. (b) Linear spectroscopy of the resonator with varying external flux. The resonance is observed to tune with the external flux $\Phi_\mathrm{e}$ in the usual way and has a maximum frequency $\omega_{\mathrm{r,max}}/2\pi = 8.31 \, \mathrm{GHz}$. (c) Close up at the frequency indicated by the black dashed line in the wider tuning plot. An anti-crossing of the microwave resonance and a mechanical mode at frequency $\omega_\mathrm{m}/2\pi = 5.9754 \, \mathrm{GHz}$ is clearly observed, with the mechanical feature only visible when the resonator is tuned in close proximity.  The  data is collected at a higher frequency resolution within a $25 \, \mathrm{MHz}$ band around the mechanical frequency in order to better resolve the mechanical mode away from resonance. The value of $\Phi_\mathrm{e}$ at which the two modes are directly on resonance is marked by a black dashed line. (d) Line cut at the resonance showing the two dips observed in the reflection spectrum. A least-squares fit (solid black lines) is overlaid with the raw data, showing close agreement with the model.}
    \label{fig:linspec}
\end{figure*}

\begin{figure}[ht]
    \centering
    \includegraphics[width=0.9\linewidth]{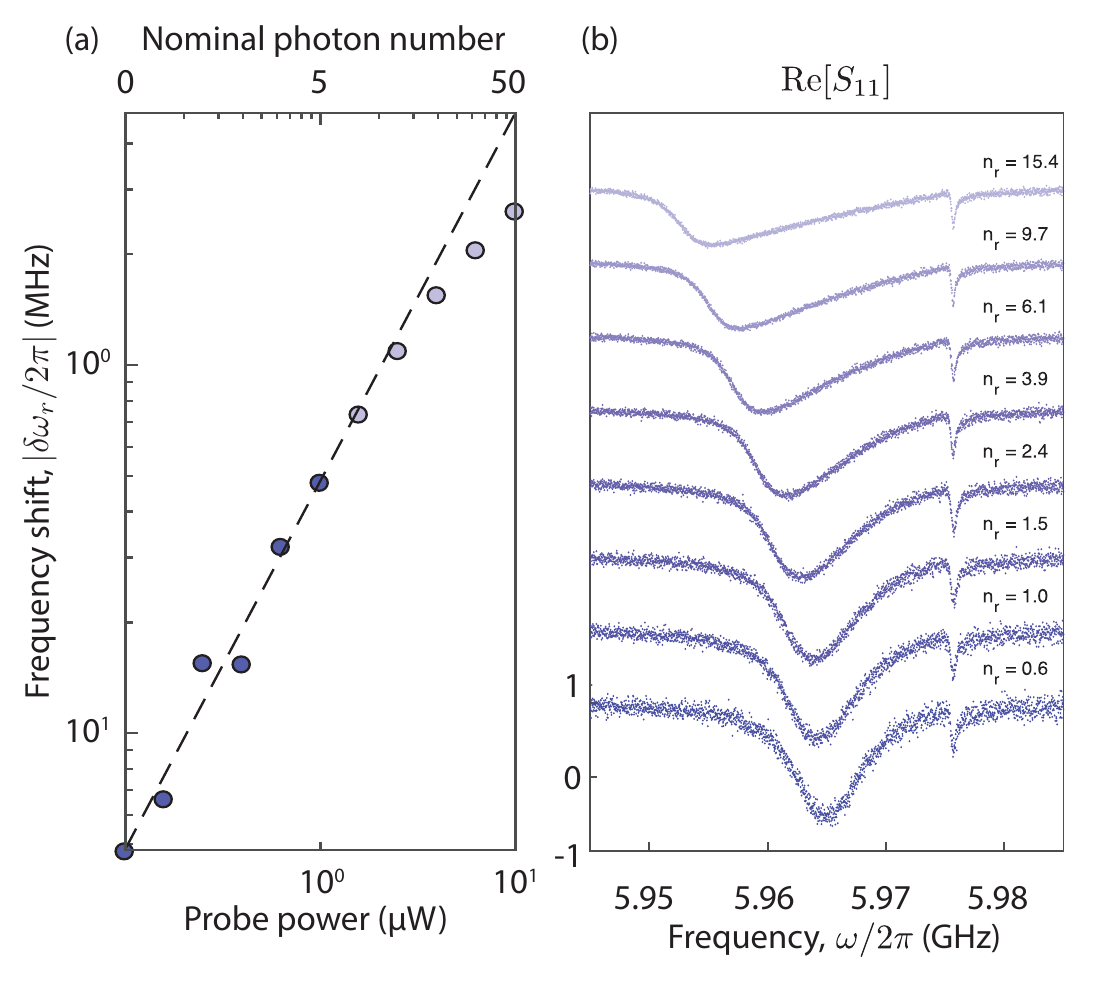}
    \caption{Nonlinear spectroscopy. (a) Frequency shift $\delta \omega_\mathrm{r} \equiv \omega_\mathrm{r}(n_\mathrm{r}) - \omega_\mathrm{r}(0)$ of the microwave resonance as a function of the probe tone power. Since the anharmonicity $\chi$ is negative the resonator redshifts as its occupation $n_\mathrm{r}$ increases; here we plot the absolute value of the shift for clarity. As expected from a linearized model of a resonator with a Kerr nonlinearity, the shift has a linear dependence for weak driving strengths (dark blue points), but deviates from this trend at stronger driving (light blue points). We can calibrate the on-resonance occupation $n_\mathrm{r}$ by fitting $\delta \omega_\mathrm{r}$ to a line in the weak driving regime (dashed black line), using a value of $\chi/2\pi = -2.0 \, \mathrm{MHz}$ for the anharmonicity. (b) Reflection spectra at various values of $n_\mathrm{r}$, with the resonator placed to the red side of the mechanical mode. As the occupation increases the resonator redshifts as expected, while the mechanical mode remains unchanged.}
    \label{fig:nonlinspec}
\end{figure}

In Fig. \ref{fig:fab}(b) we show a set of scanning-electron micrographs of a finished device nearly identical to the one used in this experiment. 
The full microwave circuit is shown in the center. The charge line (highlighted blue) is capacitively coupled to the resonator and is used for driving and readout. The flux line (highlighted red) is used to apply either DC or RF magnetic fields to the SQUID array and tune the resonator frequency. The flux line is shorted to ground in a symmetric configuration in order to reduce leakage of photons through the mutual inductance between the resonator and the line. The junction array, placed $5 \, \mu\mathrm{m}$ away from the flux line, is composed of $N_\mathrm{SQ} = 17$ nominally identical SQUIDs in series and has a total inductance $L_\mathrm{r} \approx 11 \, \mathrm{nH}$ inferred by measuring the normal-state resistance of three copies of the array on the same chip. The two terminals of the SQUID array are routed to a set of electrodes used to address six independent phononic crystal defect cavities. These electrodes, the rest of the wiring, and the immediate environment of the resonator amount to a total capacitance of $C_\mathrm{r} = 33\,\mathrm{fF}$, determined from finite-element electrostatics simulations.

Each of the six cavities has the same nominal mirror cell design and therefore the same phononic band structure. As a result the modes that are supported by the cavities  appear in the same frequency bands. In order to spectrally resolve these modes, we sweep the length $a_\mathrm{def}$ of the defect cells, from $1.4 \, \mu \mathrm{m}$ to $1.65 \, \mu \mathrm{m}$ in steps of $50 \, \mathrm{nm}$. Because the bandgap is quite small --- only a small percentage of the center frequency --- many defects do not support localized modes of the correct polarization. By scaling the defect across the six cavities, we therefore increase the likelihood of generating and observing a localized mode.

\section{Modeling and measurement results}

We model our system as a microwave-frequency electromagnetic mode with annihilation operator $\aop$ and frequency $\omega_\mathrm{r}$ that is linearly coupled, with a rate $g$, to a mechanical mode $\bop$ at frequency $\omega_\mathrm{m}$. This model is valid so long as we are interested in a range of frequencies sufficiently distant from other mechanical resonances in the system as compared to the relevant interaction rates. The Hamiltonian is 
\begin{equation}
\label{ham}
\Hop/\hbar = \omega_r \aopd \aop + \frac{\chi}{2}\hat{a}^{\dagger 2} \aop^2 + \omega_m \bopd\bop + g(\aop\bopd + \aopd\bop),
\end{equation}
where $\chi$ is the Kerr nonlinearity of the microwave mode introduced by array. For an array of $N_\mathrm{SQ}$ identical SQUIDs this is given by $\hbar \chi = -E_\mathrm{C} / N_\mathrm{SQ}^2$, where $E_\mathrm{C} = e^2/(2C_\mathrm{r})$ is the charging energy \cite{Girvin2011}. For this device $\chi/2\pi \approx -2 \, \mathrm{MHz}$, which is larger than the typical anharmonicity of parametric amplifier devices \cite{Zhou2014} but significantly smaller than that of transmon qubits \cite{Koch2007}. We can further include the effect of a coherent drive sent into the input port by adding a drive term $\Hop_\mathrm{d}/\hbar = -i\sqrt{\kappa_\mathrm{e}}(\aopd \alpha_\mathrm{in}^{-i\omega_\mathrm{d} t} - \mathrm{h.c.})$ to the Hamiltonian, where $\kappa_e$ is the extrinsic decay rate of the microwave mode into the readout channel and $\omega_\mathrm{d}$ is the drive frequency. For sufficiently weak driving the system response is linear and the Kerr term in Eq. (\ref{ham}) can be neglected. Specifically, this is valid if $\chi \langle{\aopd \aop}\rangle \ll \kappa$, when the frequency shift induced by the drive is much smaller than the total electromagnetic linewidth $\kappa$ \cite{Eichler2014}.

We perform our characterization measurements at the bottom plate of a dilution refrigerator at a temperature of $T = 7 \, \mathrm{mK}$. We probe the system by measuring the reflection spectrum $S_{11}(\omega)$ through the charge port of the device (full details of the measurement setup are provided in Appendix \ref{app:experimental_setup}). In Fig. \ref{fig:linspec}(a) we show a typical normalized reflection spectrum of the resonator  in the linear regime, in this case tuned to a frequency of $\nu = 5.9 \, \mathrm{GHz}$ far detuned from any mechanical resonance. The reflection coefficient is  $S_{11}(\omega) = -1 + 2\eta_\mathrm{e} \chi_\mathrm{r}(\omega)$, where $\eta_\mathrm{e} \equiv \kappa_\mathrm{e}/\kappa$ is the coupling efficiency and $\chi_\mathrm{r}(\omega) = \left[ 2i(\omega - \omega_\mathrm{r})/\kappa + 1\right]^{-1}$ is the dimensionless susceptibility (see Appendix \ref{app:spectra} for details). Fitting the data to this model, we obtain $\kappa/2\pi = 11 \, \mathrm{MHz}$ and $\kappa_\mathrm{e}/2\pi = 6.3 \, \mathrm{MHz}$ at a resonator frequency of 5.9 GHz. This frequency is more than $2~\text{GHz}$ detuned from the flux sweet spot, leading to an intrinsic linewidth $\kappa_\mathrm{i} = \kappa - \kappa_\mathrm{e}$ that is dominated by flux noise dephasing (see Appendix ~\ref{app:sqr_characterization}).  In Fig. \ref{fig:linspec}(b) we show the linear spectroscopy results for a range of values of the external magnetic flux $\Phi_\mathrm{e}$, illustrating the DC-bias response $\omega_\mathrm{r}(\Phi_\mathrm{e}) = \omega_{\mathrm{r,max}}\sqrt{|\cos(2\pi\Phi_\mathrm{e}/\Phi_0)|}$ of the resonator frequency. We infer $\omega_{\mathrm{r,max}}/2\pi = 8.31 \, \mathrm{GHz}$, lying outside of our measurement band. Further, since the SQUIDs are composed of nominally identical junctions the lower frequency part of our tuning curve also lies outside of the measurement band. 

We can now use the tunable response of the resonator to look for additional signatures in the spectrum. Tuning the resonance from the top of our measurement band at $\nu \approx 8~\mathrm{GHz}$ down to $\nu \approx 5~ \mathrm{GHz}$, we find a series of resonances that anti-cross with the microwave mode, largely concentrated in the $6 - 6.5~\mathrm{GHz}$ range. In Fig. \ref{fig:linspec}(c) we show the anti-crossing of the most strongly coupled mechanical mode we found for this device, along with a line cut at the point of minimum detuning shown in Fig. \ref{fig:linspec}(d). Using the entire anti-crossing dataset, we extract the parameters of the mechanical mode by fitting the spectrum to the simple linear input-output model described in Appendix~\ref{app:spectra}. In the case of a single mechanical mode coupled to the readout resonator, the reflection spectrum can be written as
\begin{equation}
\label{eq:two_mode_reflection}
S_{11}(\omega) = -1 + \frac{1}{1 + C \chi_\mathrm{m}(\omega) \chi_\mathrm{r}(\omega)}  2\eta_\mathrm{e}\chi_\mathrm{r}(\omega),
\end{equation}
where $\chi_\mathrm{m}(\omega) = \left[2i(\omega - \omega_\mathrm{m})/\gamma + 1\right]^{-1}$ is the dimensionless mechanical susceptibility and $C \equiv 4g^2 / \kappa \gamma$ is the cooperativity. A least-squares fit to this model results in $\omega_\mathrm{m}/2\pi = 5.9754 \, \mathrm{GHz}$, $g/2\pi = 1.65 \pm 0.07 \, \mathrm{MHz}$ and $\gamma/2\pi = 220 \pm 70 \, \mathrm{kHz}$, corresponding to a mechanical quality factor $Q_\mathrm{m} = \omega_m/\gamma \approx 3 \times 10^4$. The maximum cooperativity, i.e., the ratio of the mechanical resonator's electromagnetic read-out coupling to its intrinsic losses, approaches $C \approx 4.5$ on resonance. Crucially, our mode lies in a ``quiet'' region where the closest observed mechanical modes are $50~\textrm{MHz}$  and $250~\textrm{MHz}$ below and above, respectively (see Appendix~\ref{app:mech_spectrum}).

In order to better understand the measured electromechanical response, we perform finite-element simulations of the full $\mathrm{LiNbO_3}$ structure, simultaneously solving the equations of elasticity, electrostatics, and their coupling via piezoelectricity. Following a procedure described in Ref.~\cite{Arrangoiz-Arriola2016}, we numerically calculate the electromechanical admittance function $Y_\mathrm{m}(\omega)$ seen at the electrical terminals of a single phononic cavity~\cite{comsol2013} and generate an effective circuit using Foster synthesis~\cite{Nigg2012}. Using this technique we calculate coupling rates in the range $g/2\pi \approx 1.5 - 2.5 \, \mathrm{MHz}$ for the cavity geometries present in this device, in agreement with the measurement.

We measure the reflection spectra at higher drive power levels to verify the expected linearity of the mechanical resonance, and to distinguish it from other degrees of freedom, such as two-level systems (TLS) that have been observed in chip-scale devices~\cite{Neeley2008}. The strong Kerr nonlinearity of the resonator allows us to calibrate the coherently-driven photon occupation. We set the resonator frequency to $\omega_\mathrm{r}/2\pi = 5.90 \, \mathrm{GHz}$, detuned from the mechanical mode, and vary the probe power. For very low powers, we can approximate the effect of the drive as a frequency shift $\delta \omega_\mathrm{r} = \chi \langle \aopd \aop \rangle /2$ [Fig. \ref{fig:nonlinspec}(a)] and use this to extract the photon number. For low probe powers, we observe a linear dependence of the frequency shift as expected from a linearized model in which the steady-state occupation of the resonator redshifts the frequency seen by the probe tone. However, as the probe power is increased, a more complex nonlinear response is observed as evidenced by the deviation of the estimated $\delta \omega_\mathrm{r}$ from the simplified linear dependence. We use the lower power points to obtain a nominal calibrated photon number $n_\mathrm{r}=\langle \aopd \aop \rangle$, which is valid at low drive strengths and represents an upper bound to the occupation when extrapolated to stronger drives. This requires us to accurately estimate $\chi$, which we do in two different ways: first using the measured resonator frequency and the normal-state junction resistance, and second by simulating the capacitance matrix of the device. Both of these methods give us nearly the same value of $\chi/2\pi=-2.0~\textrm{MHz}$.
We now place the resonator to the red side of the mechanical mode and change the driving strength while sweeping the probe frequency to obtain the traces shown in Fig. \ref{fig:nonlinspec}(b). We observe the microwave mode broaden and redshift as the occupation is increased to a few photons, while the mechanical mode remains unchanged. We therefore conclude that the observed resonance is not due to a TLS. Additionally, we note that the frequency and linewidth of the observed resonance remained constant over several experimental runs that involved temperature cycling the device.

\section{Outlook}

We have demonstrated efficient coupling between a localized phononic cavity and a superconducting microwave circuit. The cooperativity $C \sim 4$ is already sufficient for  efficient conversion of microwave photons to highly localized microwave phonons that can in turn be up-converted efficiently to optical photons~\cite{Liang2017} -- a promising route for microwave-to-optical conversion~\cite{Safavi-Naeini2010,Bochmann2013a}. The performance of the device can be further improved by increasing the size of the bandgap to allow for higher mechanical $Q$. Larger phononic bandgaps lead to greater robustness to fabrication imperfections, which we believe currently limit the coherence time of the resonances (see Appendix \ref{app:mech_spectrum}). In addition, optimizing the electrode placement and mode profile can lead to an increase in the coupling rate $g$. 

For quantum acoustic structures to become competitive with the best electromagnetic cavities, higher interaction rates $g$ and quality factors $Q$ need to be achieved while minimizing spurious resonances to allow for fast gate operations~\cite{Pechal2018}. Interestingly, due to the small capacitance of the transducer and the ability to minimize crosstalk between resonances through phonon bandgap engineering, this architecture lends itself well to engineering systems where many bosonic linear modes couple to a single qubit~\cite{Naik2017}. Whether such a quantum acoustic approach will be competitive in the realm of quantum information processing relies on improvements in the $g$ and $Q$ of the devices, which will be the focus of  future work.

\section{Acknowledgements} 
We gratefully acknowledge R. Patel, C. Sarabalis, R. Van Laer, and N. L{\"o}rch for useful discussions. This work was supported by NSF ECCS-1509107, NSF ECCS-1708734, ONR MURI QOMAND, and start-up funds from Stanford University. ASN is grateful for support from Terman, Hellman, and Packard Fellowships. PAA and JDW are partially supported by the Stanford Graduate Fellowship (SGF), and MP is partially funded by the Swiss National Science Foundation postdoctoral fellowship. Device fabrication was performed at the Stanford Nano Shared Facilities (SNSF) and the Stanford Nanofabrication Facility (SNF). The SNSF is supported by the National Science Foundation under Grant No. ECCS-1542152. EAW was partly supported by an SNSF fellowship.

\appendix
\section{Reflection spectra}
\label{app:spectra}

We model the mechanical system as a collection of harmonic modes $\{\bop_i\}$ linearly coupled to a Kerr oscillator, which in turn is coupled to a single input/output channel for driving and readout.

The Hamiltonian of the system is 
\begin{multline}
\Hop/\hbar = \omega_\mathrm{r} \aopd \aop + \frac{\chi}{2} \hat{a}^{\dagger 2} \aop^2 + 
\sum\limits_{k} \omega_\mathrm{m}^{(k)} \bopd_k \bop_k + \\ 
\sum\limits_k g_k (\aop + \aopd)(\bop_k + \bopd_k) - i\sqrt{\kappa_\text{e}}\left(\aopd \alphain e^{-i\omega_\text{d} t} - \text{h.c.}\right) + \Hop_B,
\end{multline}
where $\omega_\mathrm{r}$ is the frequency of the microwave mode, $\chi$ is the anharmonicity, $\{\omega_\mathrm{m}^{(k)}\}$ are the frequencies of the mechanical modes, and $\{g_k\}$ are their coupling rates to the microwave mode. We have explicitly included a coherent driving field at frequency $\omega_\mathrm{d}$ (which couples to the system at rate $\kappa_\mathrm{e}$) and bundled all other bath terms into $\Hop_B$ --- following a standard input-output treatment, these terms simply generate additional decay terms in the Heisenberg equations. We can eliminate the time dependence in $\Hop$ by going into an interaction frame with respect to $\Hop_0/\hbar \equiv \omega_\mathrm{d} \left(\aopd \aop + \sum \bopd_k\bop_k \right)$. The transformed Hamiltonian (now omitting the bath terms) becomes
\begin{multline}
\label{H_rot_frame}
\Hop/\hbar = -\Delta_\text{r} \aopd \aop + \frac{\chi}{2}\hat{a}^{\dagger 2} \aop^2 - \sum\limits_{k}\Delta_\mathrm{m}^{(k)} \bopd_k \bop_k \\
+ \sum\limits_{k} g_k (\aop\bopd_k + \aopd\bop_k) - i\sqrt{\kappa_\text{e}}\left(\aopd \alphain - \text{h.c.} \right), 
\end{multline}
where $\Delta_j \equiv \omega_\mathrm{d} - \omega_j$. 

We can  neglect the nonlinear term in the weak-drive regime where $\chi|\alpha_{\mathrm{in}}|^2 \ll \kappa^2$. Our experiment only measures the average output field amplitudes in steady state, given by $\langle \aop \rangle \equiv \alpha$ and $\langle \bop_k \rangle \equiv \beta_k$. These obey the Heisenberg equations of motion 
\begin{align}
\label{eom_r_td}
\dot{\alpha} &= \left(i\Delta_\mathrm{r} - \frac{\kappa}{2}\right)\alpha - i\sum\limits_k g_k \beta_k + \sqrt{\kappa_\mathrm{e}}\alphain \\
\label{eom_m_td}
\dot{\beta}_k &= \left(i\Delta_\mathrm{m}^{(k)} - \frac{\gamma_{k}}{2}\right)\beta_k - i g_k \alpha,
\end{align}
which can be written in the Fourier domain as
\begin{align}
\frac{\kappa}{2} \chi_\mathrm{r}^{-1}(\omega) \alpha &= - i\sum\limits_k g_k \beta_k + \sqrt{\kappa_\mathrm{e}}\alphain \\
\frac{\gamma_{k}}{2} \chi_{\mathrm{m},k}^{-1}(\omega) \beta_k &= - i g_k \alpha.
\end{align}
We define the bare dimensionless susceptibilities
\begin{align}
\chi_\mathrm{r}(\omega) &= \left[-2i(\omega + \Delta_\mathrm{r})/\kappa +1\right]^{-1} \\ 
\chi_{\mathrm{m},k}(\omega) &= \left[-2i(\omega + \Delta_\mathrm{m}^{(k)})/\gamma_{k} + 1\right]^{-1},
\end{align}
where $\kappa = \kappa_\mathrm{e} + \kappa_\mathrm{i}$ and $\{\gamma_{k}\}$ are the total decay rates of the microwave and mechanical modes, respectively.  Together with the input-output boundary condition $\alphaout = -\alphain + \sqrt{\kappa_\mathrm{e}}\alpha$, we can then directly solve for the reflection coefficient $S_{11} \equiv \alphaout/\alphain$, and obtain
\begin{equation}
S_{11}(\omega) = -1 + \frac{2\kappa_\mathrm{e} / \kappa}{1 + \sum_k C_k \chi_{\mathrm{m},k}(\omega) \chi_\mathrm{r}(\omega)}  \chi_\mathrm{r}(\omega),
\end{equation}

where $C_k \equiv 4g_k^2 / \kappa \gamma_{k}$ is the cooperativity (or readout efficiency) for mode $k$. We can finally assume that a single mechanical mode $\bop_k \equiv \bop$ is relevant at a given resonator frequency $\omega_\mathrm{r}$, in the sense that it is the only mode that imprints a measurable signature in the reflection signal. Some algebraic manipulation leads us to the expression for $S_{11}(\omega)$
\begin{equation}
S_{11}(\omega) = -1 + \frac{2\kappa_\mathrm{e} / \kappa}{1 + C \chi_{\mathrm{m}}(\omega) \chi_\mathrm{r}(\omega)} \chi_\mathrm{r}(\omega),
\end{equation}
shown in the main text and used for fitting the data.

\section{Wide-band characterization of SQUID array resonator}
\label{app:sqr_characterization}

\begin{figure}[ht]
    \centering
    \includegraphics[width=\linewidth]{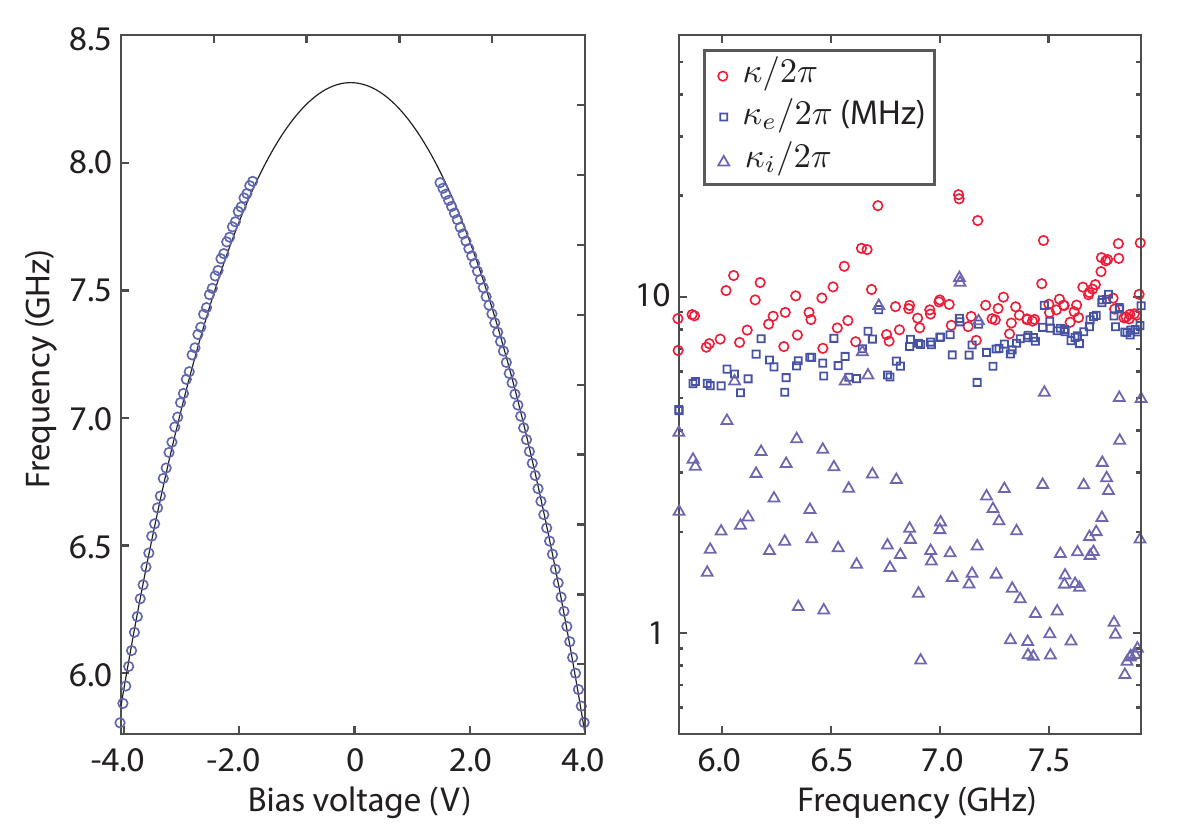}
    \caption{Wide band resonator characterization. Calibration of the external flux as a function of the applied bias voltage (left), including data points (circles) and fit to theory. The flux sweet spot at $\omega_\mathrm{r,max} = 8.31 \, \mathrm{GHz}$ lies outside our measurement band. The total, extrinsic, and intrinsic linewidths ($\kappa, \kappa_\mathrm{e}$, and $ \kappa_\mathrm{i}$, respectively) are also plotted as a function of frequency. A monotonic decrease in $\kappa_\mathrm{i}$ as the resonator is tuned towards the flux sweet spot is attributed to a reduction in the flux noise contribution to the linewidth.}
    \label{fig:sqr_calib_fig}
\end{figure}

The center frequency of the SQUID array resonator is widely tunable, allowing us to probe the the mechanical mode spectrum over a large range of frequencies. In Fig. \ref{fig:sqr_calib_fig} we plot the  resonator frequency $\omega_\mathrm{r}/2\pi$ as a function of the bias voltage applied to run a current through the on-chip flux line (see Appendix \ref{app:experimental_setup} for details) along with a fit to the function $\omega_\mathrm{r}(V) = \omega_{\mathrm{r,max}}\sqrt{|\cos(GV + \phi_\mathrm{offset})|}$, giving us a calibration of the external flux $\Phi_\mathrm{e}$ threading the SQUIDs. We infer that the flux-insensitive point is at  $\omega_{\mathrm{r,max}}/2\pi = 8.31 \, \mathrm{GHz}$ and lies outside of our measurement band. At every bias point, we fit the reflection spectrum $S_{11}(\omega)$ to the model derived in Appendix \ref{app:spectra} (Eq. \ref{eq:two_mode_reflection} with $g=0$) in order to extract the total and extrinsic resonator linewidths ($\kappa$ and $\kappa_\mathrm{e}$, respectively), also shown in Fig. \ref{fig:sqr_calib_fig}. We also define an intrinsic linewidth $\kappa_\mathrm{i} \equiv \kappa - \kappa_\mathrm{e}$, which contains contributions from both energy relaxation and pure dephasing. Since our scattering parameter measurement uses only mean field amplitudes, we do not have the ability to separate these two contributions, but we can still look at the frequency dependence of $\kappa_\mathrm{i}$ in order to gain insight into the decoherence mechanisms affecting this device. We find that $\kappa_\mathrm{e}$ increases with frequency as expected from capacitive coupling to the feedline, whereas $\kappa_\mathrm{i}$ decreases as $\omega_\mathrm{r}/2\pi$ approaches the flux-insensitive point. We attribute this dependence to a strong contribution of flux-noise dephasing to the intrinsic linewidth. In fact, we see that at the mechanical frequency $\omega_\mathrm{m}/2\pi \approx 6 \, \mathrm{GHz}$ the intrinsic linewidth is dominated by flux noise, suggesting that the coherence times in future experiments can be improved by operating the tunable circuits near or at the flux sweet spot.

\section{Complete mechanical spectrum of the system}
\label{app:mech_spectrum}

\begin{figure}[ht]
    \centering
    \includegraphics[width=\linewidth]{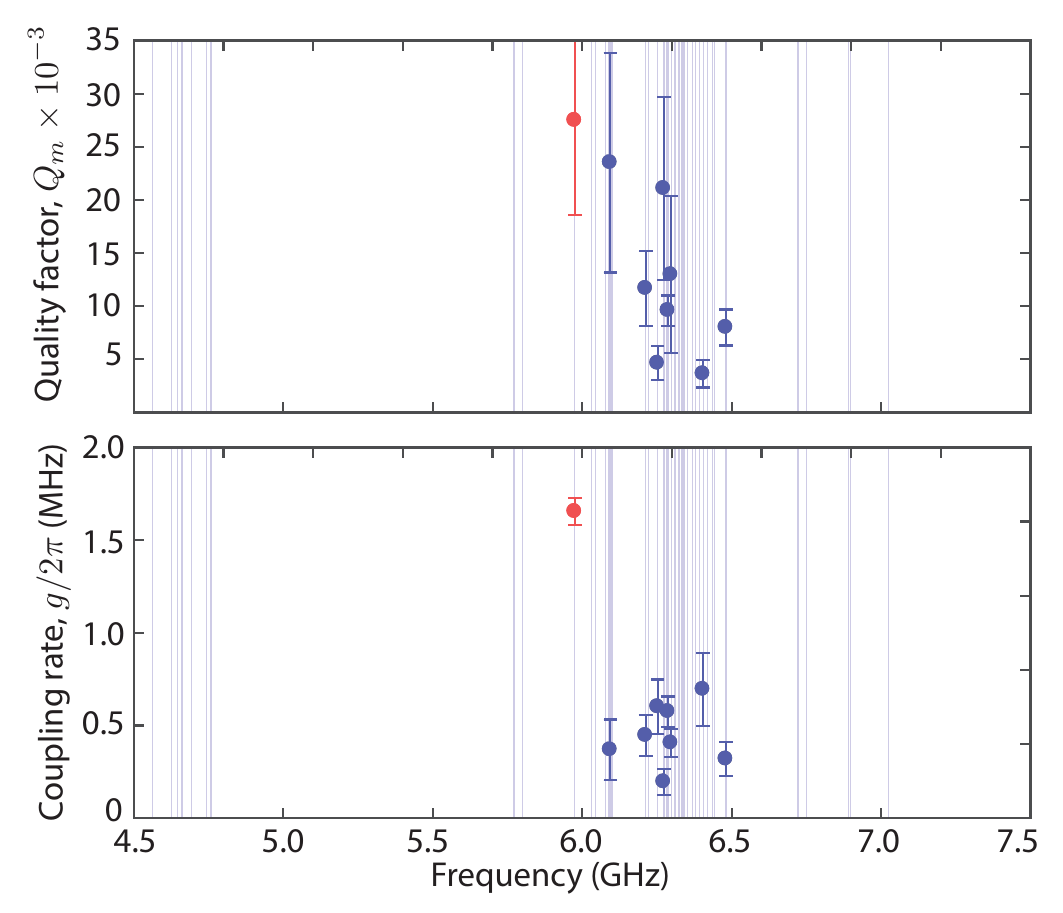}
    \caption{Mechanical spectrum. Quality factors $Q_\mathrm{m}$ (top) and coupling rates $g$ (bottom) of various modes are plotted as a function of frequency. The position of all modes observed in this device are indicated by vertical blue lines, clearly showing that the resonances tend to tightly cluster within certain regions. In both plots, the resonance at $\omega_\mathrm{m}/2\pi = 5.975 \, \mathrm{GHz}$ presented in the main text is indicated by the red point. The quality factors and coupling rates are obtained through reflection spectra collected at various detunings $\Delta = \omega_\mathrm{r} - \omega_\mathrm{m}$ around each mechanical mode and fitting them to Eq. \ref{eq:two_mode_reflection}; error bars indicate the standard deviation of the parameter estimates for these fits.}
    \label{fig:other-modes}
\end{figure}

In addition to the mode at $\omega_\mathrm{m}/2\pi = 5.9754 \, \mathrm{GHz}$ presented in the main text, we observe other modes distributed over a wide range of frequencies. In Fig. \ref{fig:other-modes} we show the complete mechanical spectrum of this device. The positions of all observed modes are indicated with vertical lines, and we plot the quality factor $Q_\mathrm{m}$ and coupling rate $g$ of nine modes with sufficiently strong signatures in $S_{11}$ to be fit reliably to the model. The mode presented in the main text is indicated in red. Interestingly, we find that \emph{all} modes have quality factors on the order of $10^4$, consistent with the hypothesis that their losses are dominated by acoustic radiation, or ``clamping" loss. These measurements are also consistent with previous studies of losses in phononic crystal cavities made from silicon that lack full phononic bandgaps~\cite{Chan2012thesis}. The size of the bandgap in this work is not very large ($<5\%$ of its center frequency according to our finite element simulations), which is of comparable magnitude to the fabrication-induced disorder in the phononic crystals. This allows trapped phonons to tunnel out of the defect region and irreversibly escape through the clamping points \cite{Chan2012thesis}. In order to suppress this loss channel, future devices will require larger bandgaps, which can be achieved through further improvements to the design and fabrication. We also observe that with the exception of mode presented in the main text --- indicated by the red point at $g/2\pi \approx 1.6 \, \mathrm{MHz}$ --- all modes have coupling rates on the order of $ 100 \, \mathrm{kHz}$. This reduction in the coupling rate has been observed in silicon optomechanical crystals as the modes are tuned outside of a bandgap region~\cite{Alegre2010}. In our case, finite element simulations of all six cavity geometries present in this device predict rates between $1.5 - 2.5 \, \mathrm{MHz}$, leading us to conclude that only one resonance in the spectrum is tightly confined to the defect in the way predicted by the simulations. The coupling rates can therefore be improved by engineering cavities with larger bandgaps, optimizing the geometry of the defect, and changing the placement of the electrodes.

\section{Experimental setup}
\label{app:experimental_setup}

\begin{figure}[ht]
    \centering
    \includegraphics[width=0.9\linewidth]{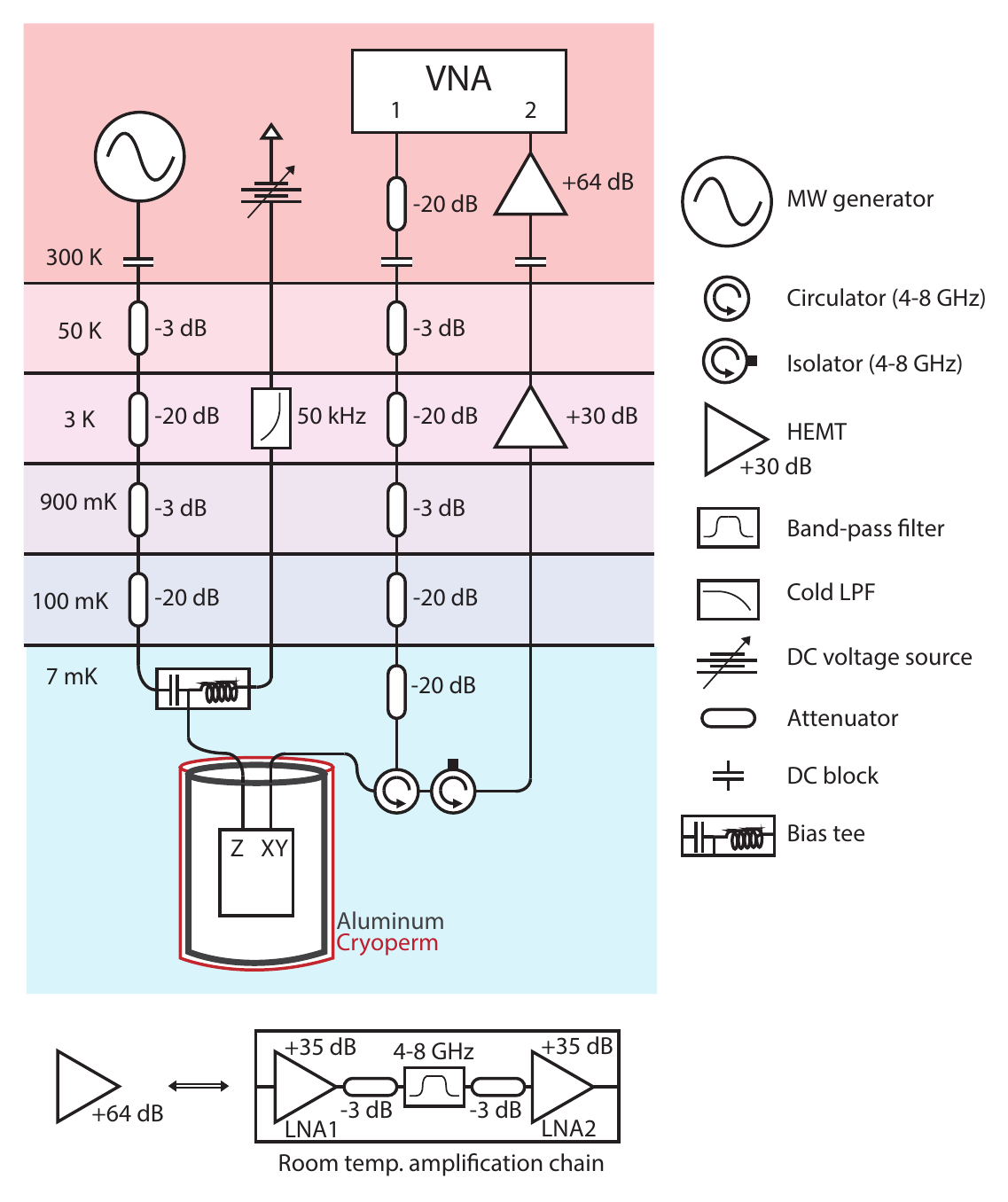}
    \caption{Experimental setup schematic. The sample is placed inside a multilayer magnetic shield (2.5 mm-thick aluminum and 1 mm-thick Cryoperm [MuShield, Inc.]) at the mixing chamber plate of a dilution refrigerator with base temperature $T \approx 7 \, \mathrm{mK}$. The measurement and control electronics for probing the scattering parameters and applying a flux bias are illustrated.}
    \label{fig:setup}
\end{figure}

Our sample is packaged in a copper enclosure to protect it from stray radiation and limit spurious modes. The package is placed inside a multi-layer magnetic shield anchored to the mixing-chamber plate ($T\approx 7\,\mathrm{mK}$) of a cryogen-free dilution refrigerator. A Rhode \& Schwartz ZNB20 vector network analyzer (VNA) generates a probe tone that is sent down to the input port of the device through a cascade of attenuators thermalized to various temperature stages of the refrigerator. A circulator (QuinStar {QCY-060400C000}) separates the input and output signals and an additional isolator (QuinStar {QCY-060400C000}) protects the device from hot ($T\sim 3\, \mathrm{K}$) radiation in the output line. The output signal is routed up to the 3 K stage through superconducting NbTi cables, where it is amplified by a high-electron mobility transistor (HEMT) amplifier (Caltech {CITCRYO1-12A}). The signal is further amplified at room temperature by two low-noise amplifiers (Miteq {AFS4-02001800-24-10P-4} \& {AFS4-00100800-14-10P-4}) with a 4-8 GHz bandpass filter ({Keenlion KBF-4/8-Q7S}) between them before being detected at the VNA. 

Flux biasing is provided by a programmable voltage source (SRS SIM928). The DC voltage passes through a cold low-pass filter (Aivon {Therma-24G}) at the 3 K stage and enters the DC port of a bias tee (Anritsu K250) mounted at the mixing-chamber plate. In addition an AC flux can be applied with a microwave generator (Keysight {E8257D}), which sends a tone to the RF port of the bias tee through an additional attenuated line, though this capability is not used in this experiment. Finally, the DC+RF output of the tee is sent directly to the flux port of the device.

\bibliographystyle{apsrev_ASN}

\end{document}